\documentclass[pra,showpacs,twocolumn]{revtex4}

\usepackage{psfrag}
\usepackage{graphicx}

\begin{document}

\title{Finite Size Corrections to Entanglement in Quantum Critical Systems}
\author{F. C. \surname{Alcaraz}}
\email{alcaraz@ifsc.usp.br}
\affiliation{Instituto de F\'{\i}sica de S\~ao Carlos, Universidade de S\~ao Paulo, \\
Caixa Postal 369, 13560-590, S\~ao Carlos, SP, Brazil.}
\author{M. S. \surname{Sarandy}}
\email{msarandy@if.uff.br}
\affiliation{Instituto de F\'{\i}sica, Universidade Federal Fluminense,
Av. Gal. Milton Tavares de Souza s/n, Gragoat\'a, 24210-346, Niter\'oi, RJ, Brazil.}
\date{\today }

\begin{abstract}
We analyze the finite size corrections to entanglement in quantum critical systems. 
By using conformal symmetry and density functional theory, we discuss the structure 
of the finite size contributions to a general measure of ground state entanglement, which 
are ruled by the central charge of the underlying conformal field theory. 
More generally, we show that all conformal towers formed by an infinite number of excited states (as 
the size of the system $L \rightarrow \infty$) exhibit a unique pattern of entanglement, which differ only at leading order $(1/L)^2$. 
In this case, entanglement is also shown to obey a universal structure, given by the anomalous dimensions of the primary 
operators of the theory. As an illustration, we discuss the behavior of pairwise entanglement for the eigenspectrum of 
the spin-1/2 XXZ chain with an arbitrary length $L$ for both periodic and twisted boundary conditions. 
\end{abstract}


\maketitle

\section{Introduction}

\label{introduction}

In recent years, the observation that entanglement may play an important role at a 
quantum phase transition~\cite{Osterloh:02,Nielsen:02,Vidal:03,Amico:08} has motivated intensive  
research on the characterization of critical phenomena via quantum information concepts. 
In this direction, conformal invariance has brought valuable information about the behavior 
of block entanglement, as measured by the von Neumann entropy, in critical models.  
Indeed, conformal field theory (CFT) has been used as a powerful tool to determine 
universal properties of entanglement. Remarkably, it was shown that the entanglement entropy obeys 
a universal logarithmic scaling law for one-dimensional critical models both at zero and finite 
temperatures~\cite{Korepin:04,Calabrese:04,Keating:05}, which is governed by the 
central charge of the associated CFT. Moreover, corrections to the entanglement entropy due 
to finite size effects have also been considered~\cite{Calabrese:04,Laflorencie:05} for periodic 
and open boundary conditions. Together with approximative methods such as renormalization 
group (see, e.g. Refs.~\cite{Refael:04,Saguia:07,Lin:07,Hur:07,Fuehringer:08}) and density functional 
theory (DFT)~\cite{Franca:08}, CFT has been settled as one of the most promising approaches for 
investigating the behavior of entanglement in many-body quantum critical systems.

In this work, we will exploit in a new perspective the impact of CFT methods for the 
evaluation of entanglement at criticality. More specifically, our approach will be based 
on the statement that finite size corrections to the ground state expectation values of 
arbitrary observables are ruled by conformal invariance. This conclusion is indeed a 
consequence of two results: (1) Finite size corrections to the energy spectrum of a 
critical theory are determined by conformal invariance~\cite{Blote:86,Affleck:86,Cardy:86}; 
(2) DFT techniques imply that, under certain conditions discussed below, general observables 
can be evaluated as a function of the first derivative of the ground state energy with 
respect to a Hamiltonian coupling parameter~\cite{Schonhammer:95,Wu:05}. We then 
simultaneously apply these two results to obtain  
the finite size corrections to  ground state entanglement in critical models. As a 
by-product, conformal invariance determines the structure of entanglement in the presence of extra symmetries 
for certain higher energy states, which are the lowest energy states in each symmetrically decoupled subspace of the 
Hilbert space. For instance, if the Hamiltonian is translationally invariant and has a $U(1)$ symmetry 
due to its commutation with the magnetization operator, we can split 
out the Hilbert space in sectors of fixed momentum and magnetization. More generally, we will also show that 
all conformal towers formed by an infinite number of excited states (as 
the size of the system $L \rightarrow \infty$) exhibit a unique pattern of entanglement, which differ only at leading order $(1/L)^2$. 
This will be based on a generalization of the HK theorem for individual states belonging to conformal towers 
of critical systems. 
Finite size corrections to entanglement in these excited states will 
obey a universal structure, given by the anomalous dimensions of the primary operators of the theory.

Since our approach is applicable for any entanglement measure, it allows in particular for the 
investigation of the universality properties of pairwise entanglement measures, e.g., concurrence~\cite{Wootters:98} and 
negativity~\cite{Vidal:02}. For pairwise measures, 
criticality was first noticed through a divergence in the derivative of entanglement, 
signaling a second-order phase transition~\cite{Osterloh:02}. For first-order phase transitions, 
jumps in entanglement itself indicates quantum critical points~\cite{Bose:02,Alcaraz:04}. 
A general explanation for this distinct usual behavior of first-order and second-order phase 
transitions has been provided in Refs.~\cite{Wu:04,Wu:05} (for an explicit discussion of examples 
which do not obey this expected behavior, see Ref.~\cite{Yang:05}). From the point of view of CFT, 
we will be able to explicitly work out the finite size corrections to pairwise entanglement measures 
and show how these corrections involve universal quantities, such as the central charge or the anomalous 
dimension of primary operators associated with the CFT. As an illustration, we will consider the 
spin-1/2 XXZ chain, where an analytical expression, valid up to $o(L^{-2})$, will be provided for the negativity of 
nearest neighboring spins as a function of the size $L$ of the chain. 

\section {Energy spectrum and finite size effects in critical quantum systems}

\label{enerCFT}

Let us consider a critical theory in a strip of finite width $L$ with periodic boundary conditions. 
The transfer matrix of the theory is written as $T=\exp (aH)$, where $a$ denotes the lattice spacing and $H$ 
is the Hamiltonian. Then, for large $L$, the ground state energy density $\varepsilon(L)=E_0(L)/L$ of $H$ 
is provided by conformal invariance~\cite{Blote:86,Affleck:86}, reading
\begin{equation}
\varepsilon(L)=\varepsilon_\infty-\frac{\pi\,c\,\xi}{6}L^{-2}+o(L^{-2}),
\label{energy}
\end{equation}
where $\varepsilon_\infty$ is the energy density in the limit $L\rightarrow \infty$ and $o(L^{-2})$ 
denotes terms of any order higher than $L^{-2}$. In Eq.~(\ref{energy}), $c$ is the central 
charge of the Virasoro algebra (the conformal anomaly) and the parameter $\xi$ must be 
fixed in such a way that the equations of motion of the theory are conformally invariant~\cite{Gehlen:86}. 
The structure of the higher energy states is determined by the primary operators of the theory~\cite{Cardy:86}. 
For each operator $O_{\alpha}$ with anomalous dimension $x_\alpha$, there corresponds a tower of states with 
energy densities $\varepsilon^{\alpha}_{j,j^\prime}(L)$ given by
\begin{equation}
\varepsilon^{\alpha}_{j,j^\prime}(L)=\varepsilon(L)+2\pi\,\xi(x_\alpha+j+j^\prime) L^{-2}+o(L^{-2}),
\label{energy_ex}
\end{equation}
where $j,j^\prime=0,1,...$ are indices labelling the tower of states associated with the 
anomalous dimensions $x_\alpha$. 
Higher-order corrections to Eqs.~(\ref{energy}) and (\ref{energy_ex}) as well as convenient 
generalizations for more general boundary conditions, e.g., twisted boundary conditions, may also 
be obtained~\cite{Alcaraz:87,Alcaraz:88}. 

\section{Hohenberg-Kohn Theorem and expectation values of observables}

\label{HK}

Let us turn now to the discussion on how DFT can be allied with 
conformal invariance to extract information about expectation values of observables from 
the energy spectrum. DFT~\cite{Hohenberg:64,Kohn:65} is originally based on the Hohenberg-Kohn 
(HK) theorem~\cite{Hohenberg:64} which, for a many-electron system, establishes 
that the dependence of the physical quantities on the external potential $v(\mathbf{r})$ 
can be replaced by a dependence on the particle density $n(\mathbf{r})$. The HK theorem 
can be extended for the context of a generic quantum Hamiltonian $H$ on a lattice (see, e.g., Refs.~\cite{Schonhammer:95,Wu:05}).
In order to be specific, let us consider a quantum spin chain of size $L$ governed by 
the Hamiltonian
 \begin{equation}
H = H_{0} + \lambda \sum_{i=1}^{L} A_i,
\label{H}
\end{equation}%
where $\lambda$ is a control parameter associated with the Hermitian operators 
$A_i$ which act on the site $i$, e.g., an observable relevant to driving a quantum phase 
transition. Let us take, for simplicity, a translationally invariant chain (e.g., by assuming periodic boundary conditions). 
Then, by taking the expectation value of Eq.~(\ref{H}), we obtain 
\begin{equation}
\langle H \rangle = \langle H_0 \rangle + \lambda L \langle A \rangle ,
\end{equation}
where $\langle A \rangle \equiv \langle A_i \rangle = \langle A_j \rangle \,\,\, (\forall i,j)$ 
due to translation symmetry. Therefore
\begin{equation}
\varepsilon = \varepsilon_0 + \lambda \langle A \rangle ,
\end{equation}
where $\varepsilon = \langle H \rangle/L$ and $\varepsilon_0 = \langle H_0 \rangle/L$ are the 
energy densities associated with $H$ and $H_0$, respectively.
For a general Hamiltonian such as given in Eq.~(\ref{H}), the HK theorem can be generalized to the statement that there is a duality (in the sense of a Legendre transform) between the expectation 
value $\langle A \rangle$ (corresponding to $n(\mathbf{r})$) and the control parameter $\lambda$ (corresponding to $v(\mathbf{r})$)~\cite{Schonhammer:95,Wu:05}. In order to specify the conditions supporting this duality let us separately consider the cases of nondegenerate and degenerate Hamiltonians.

\subsection{Nondegenerate case} 
Let $\lambda$ and $\lambda^\prime$ be two fixed values of the coupling parameter in Eq.~(\ref{H}), 
which correspond to nondegenerate ground states given by 
$|\psi\rangle$ and $|\psi^\prime\rangle$, respectively. We assume that, for 
$\lambda \ne \lambda^\prime$, we have that $|\psi\rangle \ne \alpha |\psi^\prime\rangle$, 
with $\alpha$ a complex phase. This assumption means that different values of the coupling 
parameter are associated with distinct ground states. It reflects the requirement of the  
uniqueness of the potential (see, e.g., Ref.~\cite{Capelle:01}). 
A general condition to ensure the uniqueness 
of the potential for Hamiltonian~(\ref{H}) will be derived below. Then, by assuming a unique 
potential and taking two different couplings $\lambda$ and $\lambda^\prime$, the Rayleigh-Ritz variational principle allows us to write
\begin{eqnarray}
\langle \psi | H | \psi \rangle <  \langle\psi^\prime | H | \psi^\prime \rangle  
= \langle\psi^\prime | H^\prime | \psi^\prime \rangle + \left(\lambda - \lambda^\prime\right) 
L \langle A \rangle^\prime \, ,
\end{eqnarray}
where $\langle A \rangle^\prime = \langle\psi^\prime | A | \psi^\prime \rangle$ and $H$ and $H^\prime$ are the Hamiltonians associted 
with $\lambda$ and $\lambda^\prime$, respectively. Therefore
\begin{equation}
\varepsilon - \varepsilon^\prime <   \left(\lambda - \lambda^\prime\right) \langle A \rangle^\prime .
\label{rr1}
\end{equation}
Analogously, application of the variational principle for the ground state $|\psi^\prime\rangle$ 
results into
\begin{equation}
\varepsilon^\prime - \varepsilon <   \left(\lambda^\prime - \lambda\right) \langle A \rangle .
\label{rr2}
\end{equation}
By adding Eqs.~(\ref{rr1}) and~(\ref{rr2}) we obtain
\begin{equation}
\langle A \rangle^\prime \ne \langle A \rangle. 
\label{hknd}
\end{equation}
Eq.~(\ref{hknd}) expresses the HK theorem for nondegenerate ground states, stating that 
distinct densities are associated with distinct potentials. In other words, we can establish the 
map
\begin{equation}
\lambda \Longleftrightarrow |\psi\rangle \Longleftrightarrow \langle A \rangle
 = \langle\psi | A | \psi \rangle \, .
\end{equation}

\subsection{Degenerate case} 
In order to establish the HK theorem for degenerate ground states, let us consider two fixed values 
of the coupling constant, each of them associated with arbitrarily degenerate ground states:
\begin{eqnarray}
\lambda &\longleftrightarrow& q-{\textrm{degenerate ground states:}} \left\{|\psi_1\rangle,\ldots, 
|\psi_q\rangle \right\} \, , \nonumber \\
\lambda^\prime &\longleftrightarrow& q^\prime-{\textrm{degenerate ground states:}} 
\left\{|\psi^\prime_1\rangle,\ldots,|\psi^\prime_{q^\prime}\rangle \right\} . \nonumber
\end{eqnarray}
Considering that any of the ground states are equally likely, we can describe them 
by the uniformly distributed density matrices 
\begin{equation}
\rho = \frac{1}{q} \sum_{i=1}^{q} |\psi_i\rangle \langle \psi_i| \,\,\,\, , \,\,\,\,  
\rho^\prime = \frac{1}{q} \sum_{i=1}^{q^\prime} |\psi^\prime_i\rangle \langle \psi^\prime_i| \, .
\end{equation}
The requirement of uniqueness of the potential yields in the degenerate case the condition that $\rho$ and $\rho^\prime$ are 
distinct. Applying the variational principle, we obtain
\begin{equation}
{\textrm{Tr}} \left( \rho H \right) <  {\textrm{Tr}} \left( \rho^\prime H \right) 
= {\textrm{Tr}} \left( \rho^\prime H^\prime \right) + \left(\lambda - \lambda^\prime\right) 
L \langle A \rangle^\prime \, ,
\label{fd}
\end{equation}
where, here, $\langle A \rangle^\prime =  {\textrm{Tr}} \left( \rho^\prime A \right)$. 
Eq.~(\ref{fd}) implies that 
$\varepsilon - \varepsilon^\prime <   \left(\lambda - \lambda^\prime\right) \langle A \rangle^\prime$. Therefore, as before, we use the complementary equation 
$\varepsilon^\prime - \varepsilon <   \left(\lambda^\prime - \lambda\right) \langle A \rangle$
and obtain $\langle A \rangle^\prime \ne \langle A \rangle$. 
The HK map in this case can be written as
\begin{equation}
\lambda \Longleftrightarrow \rho  \Longleftrightarrow \langle A \rangle 
= {\textrm{Tr}} \left( \rho A \right) \, .
\end{equation}

\subsection{Uniqueness of the potential} 

As discussed above, the condition for the uniqueness of the potential, which is 
fundamental for the derivation of the HK theorem, is defined by the requirement that 
different values of the coupling parameter $\lambda$ are associated with distinct ground 
states of the Hamiltonian $H$. Here we will show that {\it{a necessary and sufficient condition for which 
different values of $\lambda$ are associated with distinct eigenstates of $H$ 
 is that the operators $H_0$ and $\sum_i A_i$, as given in 
Eq.~(\ref{H}), do not have common eigenstates}}. {\underline{Sufficiency}}: Suppose that two distinct couplings 
$\lambda$ and $\lambda^\prime$ yield the same eigenstate of $H$ 
\begin{eqnarray}
\left( H_0 + \lambda \sum_i A_i \right) |\psi\rangle &=& E(\lambda) |\psi\rangle \label{nu1} \\
\left( H_0 + \lambda^\prime \sum_i A_i \right) |\psi\rangle &=& E(\lambda^\prime) |\psi\rangle. 
\label{nu2}
\end{eqnarray}
Then, from Eqs.~(\ref{nu1}) and~(\ref{nu2}), we obtain 
\begin{equation}
\sum_i A_i |\psi\rangle = \frac{E(\lambda)-E(\lambda^\prime)}{\lambda - \lambda^\prime} |\psi\rangle.
\end{equation}
Therefore, in this case, $|\psi\rangle$ is also an eigenstate of $\sum_i A_i$ (as well as an eigenstate of $H_0$). 
Hence, the condition that 
$H_0$ and $\sum_i A_i$ do not have common eigenstates is sufficient for ensuring the uniqueness 
of the potential. 
{\underline{Necessity}}: Let us suppose that $H_0$ and $\sum_i A_i$ have a common eigenstate
\begin{eqnarray}
H_0 |\psi\rangle &=& E_0 |\psi\rangle \label{nu3} \\
\sum_i A_i |\psi\rangle &=& a |\psi\rangle \label{nu4} 
\end{eqnarray}
Then we obtain that $ ( H_0 + \lambda \sum_i A_i ) |\psi\rangle = (E_0 + \lambda a )|\psi\rangle$. 
Hence, by varying $\lambda$, we only change the eigenvalue (keeping the same eigenstate), which means 
that distinct couplings will lead to the same eigenstate of $H$. Therefore, the condition that 
$H_0$ and $\sum_i A_i$ do not exhibit a common eigenstate is also necessary for the uniqueness of the potential. 
In conclusion, the sufficient and necessary condition for the uniqueness of the potential can be translated by the 
noncommutation relation
\begin{equation}
\left[ H_0 ,  \sum_i A_i \right] \ne 0
\label{H0AComm}
\end{equation}
Naturally, we disregard in Eq.~(\ref{H0AComm}) the rather nonusual situation where   
$H_0$ and $\sum_i A_i$ are noncommuting observables, but $[ H_0 ,  \sum_i A_i ] |\psi\rangle$ 
results in a vanishing quantum state. 

\subsection{HK theorem for conformal towers in quantum critical models} 

Since the HK theorem is based on a variational principle, we cannot guarantee that the expectation values of the observables  
in individual excited states are in general a function of the derivative of the energy of the excited state. Naturally, as previously 
mentioned in Sec.~\ref{introduction}, the HK theorem can be applicable in the presence of symmetries to excited states that are the 
minimum energy states in a given symmetric subspace of Hilbert space. In this work, we show that, under certain conditions, the HK theorem 
can also be extended for all the individual states of conformal towers in quantum critical models. 
We begin by supposing a periodic chain 
governed by a Hamiltonian given by Eq.~(\ref{H}) which is conformally invariant in a critical interval $\lambda_{c_1} \le \lambda \le \lambda_{c_2}$. Moreover we will assume the condition (\ref{H0AComm}) for the uniqueness of the potential. 
Let us denote by $\{|\psi^{\alpha}_{j,j^\prime;d}(\lambda)\rangle\}$ the set of eigenstates associated with the energy 
$\varepsilon^{\alpha}_{j,j^\prime}(\lambda)$, with $d=1,\ldots,D$ labelling the $D$-fold degeneracy (see Eq.~(\ref{energy_ex})). 
We take the system in a uniformly distributed density matrix
\begin{equation}
\rho^{\alpha}_{j,j^\prime}(\lambda) = \frac{1}{D} \sum_{d=1}^{D} |\psi^{\alpha}_{j,j^\prime;d}(\lambda)\rangle \langle \psi^{\alpha}_{j,j^\prime;d}(\lambda)| \,.
\end{equation}
Our aim is to show that the potential $\lambda$ uniquely specifies the density 
\begin{equation}
\langle A \rangle^{\alpha}_{j,j^\prime} =  {\textrm{Tr}} \left[ \rho^{\alpha}_{j,j^\prime}(\lambda) A \right] = 
\frac{\partial \varepsilon^{\alpha}_{j,j^\prime}}{\partial \lambda} . 
\end{equation}
Therefore, the derivative 
$\partial \varepsilon^{\alpha}_{j,j^\prime}/\partial \lambda$ must be a monotonic function of $\lambda$. In order for this to occur, 
it is sufficient that: (i) $\partial \varepsilon^{\alpha}_{j,j^\prime} / \partial \lambda$ is continuous 
in the interval $\lambda_{c_1} \le \lambda \le \lambda_{c_2}$   
and (ii) $\partial^2 \varepsilon^{\alpha}_{j,j^\prime} / \partial \lambda^2 \ne 0$. Condition (i) is usually achieved for a smooth (well-behaved) 
energy. Concerning condition (ii), let us take the derivative of Eq.~(\ref{energy_ex}), which yields
\begin{equation}
\frac{\partial^2 \varepsilon^{\alpha}_{j,j^\prime}}{\partial \lambda^2} = 
\frac{\partial^2 \varepsilon}{\partial \lambda^2}  + \frac{2 \pi}{L^2} \frac{\partial^2}{\partial \lambda^2}\left[\xi(x_\alpha+j+j^\prime)\right]
+o(L^{-2}).
\label{deriv2-en}
\end{equation}
The first term in the r.h.s. of Eq.~(\ref{deriv2-en}) concerns the second derivative of the ground state energy with respect to $\lambda$. 
We can show that this term is strictly negative. Indeed, from Eqs.~(\ref{rr1}) and~(\ref{rr2}), which hold for both degenerate and non-degenerate 
ground states, we obtain
\begin{equation}
\frac{\partial}{\partial\lambda} \langle A \rangle = \frac{\partial^2 \varepsilon}{\partial \lambda^2} < 0 \, , 
\end{equation}
where $\langle A \rangle$ denotes the expectation value of $A$ taken in the ground state. Concerning the second term in the r.h.s. of 
Eq.~(\ref{deriv2-en}), it is negligible for large $L$. Consequently, we can write
\begin{equation}
\frac{\partial^2 \varepsilon^{\alpha}_{j,j^\prime}}{\partial \lambda^2} \approx 
\frac{\partial^2 \varepsilon}{\partial \lambda^2} < 0 \,\,\,\,\,\, ({\textrm{large}}\,\,\, L).
\end{equation}
Hence, $\partial^2 \varepsilon^{\alpha}_{j,j^\prime} / \partial \lambda^2$ is non-vanishing and then 
the derivative $\partial \varepsilon^{\alpha}_{j,j^\prime}/\partial \lambda$ is monotonically related to $\lambda$. 
Therefore, a $D$-fold degenerate (up to order $L^{-2}$) eigenlevel given by $\alpha$, $j$, and $j^\prime$ defines a density matrix  
$\rho^{\alpha}_{j,j^\prime}$ that can be taken either as a function of $\lambda$ or $\langle A \rangle^{\alpha}_{j,j^\prime}$. This provides an 
extension of the HK theorem for arbitrary individual eigenstates belonging to conformal towers in quantum critical models.

\section{Finite size corrections to entanglement in conformal invariant models} 

\label{ecft}

The HK theorem implies a duality between the potential $\lambda$ and the density $\langle A \rangle$. 
This behavior was revealed specially useful for the investigation of entanglement in the 
ground state of quantum systems undergoing quantum phase transitions~\cite{Wu:05}. In particular, the dependence of an arbitrary entanglement measure $M$ on the parameter $\lambda$ can be replaced by the dependence on the ground state expectation value $\langle A \rangle$~\cite{Wu:05}, which means that 
\begin{equation}
M=M(\lambda)= M(\langle A \rangle) = M(\frac{\partial \varepsilon}{\partial \lambda}) , 
\label{entanglement}
\end{equation}
where the Hellmann-Feynman theorem~\cite{Hellmann:37,Feynman:39} has been used in the last 
equality above. As discussed in the last Section, in the case of critical models, the HK theorem can also be applied 
to any state of conformal towers, which allows us to write the entanglement of such states as 
\begin{equation}
M^{\alpha}_{j,j^\prime} = M^{\alpha}_{j,j^\prime}(\lambda) = M^{\alpha}_{j,j^\prime}(\langle A \rangle^{\alpha}_{j,j^\prime})\, .
\label{M-en-ex}
\end{equation}
Eq.~(\ref{M-en-ex}) can be rewritten by observing that
\begin{equation}
\langle A \rangle^{\alpha}_{j,j^\prime} = \frac{\partial \varepsilon^{\alpha}_{j,j^\prime}}{\partial \lambda} = 
\langle A \rangle + \frac{2 \pi}{L^2} \frac{\partial}{\partial \lambda}\left[\xi(x_\alpha+j+j^\prime)\right]
+o(L^{-2})
\label{A-en-ex}
\end{equation}
By inserting Eq.~(\ref{A-en-ex}) into Eq.~(\ref{M-en-ex}) and performing a series expansion, we obtain
\begin{eqnarray}
M^{\alpha}_{j,j^\prime} &=&  
M(\langle A \rangle) + \frac{2 \pi}{L^2}   
\frac{\partial}{\partial \lambda}\left[\xi(x_\alpha+j+j^\prime)\right]
\left[\frac{\partial M^{\alpha}_{j,j^\prime}}{\partial \lambda}\right]_{\lambda=\langle A \rangle} \nonumber \\ 
&+& o(L^{-2})
\label{mcft}
\end{eqnarray} 
This means that the entanglement corresponding to all the eigenstates in the conformal towers are nearly the same as that of the 
ground state, with corrections of order $L^{-2}$. Moreover, it follows that finite size effects for an arbitrary measure of entanglement 
are ruled by conformal invariance. 

In order to evaluate entanglement at a point $\lambda=\lambda_c$, we should be able to perform a derivative 
of the energy with respect to $\lambda$ taken at $\lambda_c$. Therefore, note that we will have that Eqs.~(\ref{energy}) and 
(\ref{energy_ex}) are the starting point to determine finite size corrections to entanglement 
if and only if the theory is critical in an interval around $\lambda_c$. 
For the case of a single critical point (e.g., Ising spin-1/2 chain in a transverse field) instead of a 
critical region (e.g, XXZ spin-1/2 chain in the anisotropy interval $-1<\Delta<1$), more general expressions 
for the energy should be used, which take into account a mass spectrum (see, e.g., Ref.~\cite{Sagdeev:89}).

\section{Entanglement in the finite size spin-1/2 XXZ chain} 

As an illustration of the previous results, let us consider the spin-1/2 XXZ 
chain, whose Hamiltonian is given by
\begin{equation}
H_{XXZ}=-\frac{J}{2} \sum_{i=1}^{L} \left( \sigma^x_i \sigma^x_{i+1} + 
\sigma^y_i \sigma^y_{i+1} + \Delta \sigma^z_i \sigma^z_{i+1} \right),
\label{HXXZ}
\end{equation}
where periodic boundary conditions (PBC) are assumed. We will set the energy scale such that $J=1$. 
Entanglement for spin pairs can be quantified by the negativity~\cite{Vidal:02}, which is 
defined by 
\begin{equation}
\mathcal{N}(\rho ^{ij})=2\,\max (0,-\min_{\alpha }(\mu _{\alpha }^{ij})),
\end{equation}
where $\mu _{\alpha }^{ij}$ are the eigenvalues of the partial transpose $
\rho ^{ij,T_{A}}$ of the density operator $\rho ^{ij}$, defined as $
\left\langle \alpha \beta \right\vert \rho ^{T_{A}}\left\vert \gamma \delta
\right\rangle =\left\langle \gamma \beta \right\vert \rho \left\vert \alpha
\delta \right\rangle $. For the XXZ model, $U(1)$ invariance 
($\left[H,\sum_i \sigma_z^i\right]=0$) and translation  invariance
ensure that the reduced density matrix for spins $i$ and $j$ 
reads
\begin{equation}
\rho^{ij} = \left( 
\begin{array}{cccc}
a^{ij} & 0 & 0 & 0 \\ 
0 & b^{ij} & z^{ij}  & 0 \\ 
0 & z^{ij*}  & b^{ij} & 0 \\ 
0 & 0 & 0 & d^{ij}%
\end{array}%
\right) ,  \label{rij}
\end{equation}
where 
\begin{eqnarray}
a^{ij} &=& \frac{1}{4} \left(1+2\, G_{z}+G^{ij}_{zz}\right) \, , \nonumber \\
b^{ij} &=& \frac{1}{4} \left(1-G^{ij}_{zz}\right) \, , \nonumber \\
d^{ij} &=& \frac{1}{4} \left(1-2\,G_{z}+G^{ij}_{zz}\right) \, , \nonumber \\
z^{ij} &=& \frac{1}{4} \left[ \left(G^{ij}_{xx}+G^{ij}_{yy} \right) + i\left(G^{ij}_{xy}-G^{ij}_{yx} \right)\right], 
\label{relem}
\end{eqnarray}
where $G_{z} = \langle \sigma_z^i \rangle$ is the magnetization density (computed for any 
site $i$) and $G^{ij}_{\alpha\beta}=\langle \sigma_\alpha^i \sigma_\beta^{j} \rangle$ 
($\alpha,\beta=x,y,z$), with the expectation value taken over 
an arbitrary quantum state of the system. Moreover, invariance of $H_{XXZ}$ under 
the discrete transformations $\sigma_x \rightarrow - \sigma_x$, $\sigma_y \rightarrow \sigma_y$, 
and $\sigma_z \rightarrow - \sigma_z$ implies that $G^{ij}_{xy}=0$ and $G^{ij}_{yx}=0$. 
Therefore, the element $z^{ij}$ in Eq.~(\ref{relem}) is real, namely, 
$z^{ij}=z^{ij*}=1/4 \left( G^{ij}_{xx} + G^{ij}_{yy} \right)$. 
Then, evaluation of the negativity for spins $i$ and $j$ from 
Eq.~(\ref{rij}) yields
\begin{equation}
{\cal N}(L)=\frac{1}{2}\max\left(0,\sqrt{4 G^{2}_{z} + \left|G^{ij}_{xx}+G^{ij}_{yy}\right|^2} - G^{ij}_{zz} - 1 \right).
\label{negXXZ_gen}
\end{equation}
From now on, we will be interested in computing the negativity for nearest neighbor spins. 
The generalized HK theorem discussed in Section~\ref{HK} implies that we 
can consider $\Delta$ as the external potential and 
$\langle\sigma^z_i \sigma^z_{i+1}\rangle$ (for any site $i$) as the relevant density. Thus, 
${\cal N}(L)$ can be written as a function of $\partial \varepsilon/\partial \Delta$ for the 
ground state as well as for any minimum energy state in a sector of magnetization $m$ ($m=0,\pm 2, \ldots, \pm L$) and 
momentum $P = (2 \pi /L) p$ ($p=0,1,\ldots,L-1$). 
In this direction, it is convenient to write the correlation functions $G^{i,i+1}_{\alpha\beta}$ 
in terms of $\partial \varepsilon/\partial \Delta$, which results into
\begin{eqnarray}
G^{i,i+1}_{zz} &=& -2 \frac{\partial \varepsilon}{\partial \Delta} \, , \nonumber \\
G^{i,,i+1}_{xx}+G^{i,i+1}_{yy} &=& -2 \left(\varepsilon - \Delta \frac{\partial 
\varepsilon}{\partial \Delta} \right).
\label{correl}
\end{eqnarray}

\subsection{Ground state entanglement}

\begin{figure}[ht]
\centering {\includegraphics[angle=0,scale=0.36]{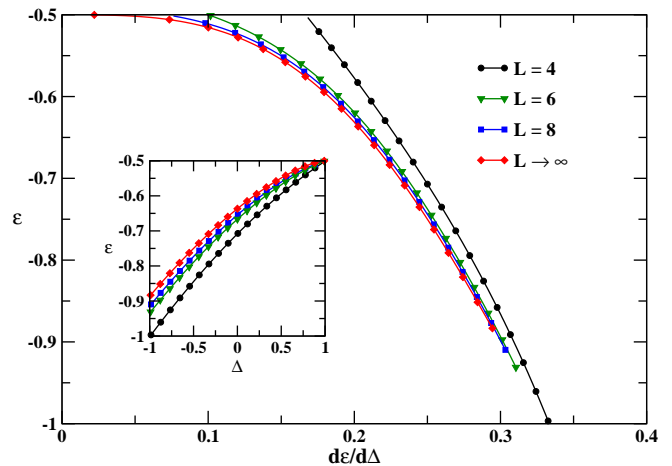}}
\caption{(color online) Density energy $\varepsilon$ as a function of $\partial \varepsilon / \partial \Delta$ 
as given by the solution of the Bethe equations for finite size chains as well as in the thermodynamic 
limit (values plotted in the range $-1 < \Delta < 1$). For finite chains with lattice sizes $L > 8$,  
the curves get nearly superposed with the curve for the infinite chain. 
Inset: Density energy $\varepsilon$ versus anisotropy parameter $\Delta$. Note that $\varepsilon$ 
can be taken either as a function of $\Delta$ (the potential) or $\partial \varepsilon / \partial \Delta$ (the density).}
\label{f1}
\end{figure}

For the ground state, we have that $G_{z}=0$. Then, by using Eq.~(\ref{correl}) and 
Eq.~(\ref{negXXZ_gen}), negativity reads
\begin{equation}
{\cal N}(L)=-\varepsilon(L)+(\Delta+1)\frac{\partial\varepsilon(L)}{\partial\Delta}
-\frac{1}{2},
\label{negXXZ}
\end{equation}
where we have used that $|G^{i,i+1}_{zz}| \le 1$ and 
$G^{i,i+1}_{xx}+G^{i,i+1}_{yy} \ge 0$ (Marshall-Peierls rule). Note that, in Eq.~(\ref{negXXZ}), 
the energy density $\varepsilon(L)$ can be seen as a function of $\partial \varepsilon/\partial \Delta$ by the HK theorem, 
which is explicitly shown in Fig.~\ref{f1}. Indeed, this implies that the negativity can be taken as a function of 
$\partial \varepsilon/\partial \Delta$, which illustrates the duality between potential and density established in Eq.~(\ref{entanglement}) 
for entanglement measures.
The XXZ model is critical in the interval $-1\le \Delta<1$, with central charge $c=1$. Then, from Eq.~(\ref{negXXZ}), we can 
determine an approximate analytical expression for the negativity in terms of energy as given by Eq.~(\ref{energy}). 
The parameter $\xi$ appearing in Eq.~(\ref{energy}) can be obtained analytically~\cite{Hamer:86} for the XXZ chain, reading 
\begin{equation}
\xi=\pi \, \frac{\sin \gamma}{\gamma}, 
\end{equation}
where $\gamma$ is defined by 
\begin{equation}
\Delta=-\cos\gamma \, , \,\,\,\,\,\gamma \in [0,\pi)
\label{gamadef}
\end{equation}
Then, substitution of Eq.~(\ref{energy}) into Eq.~(\ref{negXXZ}) yields
\begin{eqnarray}
{\cal N}_{CFT}(L)&=&{\cal N}_\infty+\frac{\pi^2c}{6\gamma L^2}\left(\sin{\gamma} 
+\frac{1+\Delta}{\gamma} \right. 
\left.+\frac{\Delta\sqrt{1+\Delta}}{\sqrt{1-\Delta}} \right) \nonumber \\ 
&+& o(L^{-2}),
\label{negXXZci}
\end{eqnarray}
where ${\cal N}_\infty$ can be computed from 
Eq.~(\ref{negXXZ}), with $\varepsilon_\infty$ and $\partial \varepsilon_\infty / \partial \Delta$ 
directly given by the solution of the model at the thermodynamic limit~\cite{Yang:66}. 
An exact value for the 
negativity ${\cal N}(L)$ can be obtained from Eq.~(\ref{negXXZ}) by computing $\varepsilon (L)$ and $\partial \varepsilon(L) / \partial \Delta$ 
via Bethe ansatz equations for each length $L$. Naturally, this amounts to a much harder computational effort for a general $\Delta$, while 
Eq.~(\ref{negXXZci}) directly provides the negativity for a finite chain up to order $L^{-2}$ with no need of solving the Bethe ansatz 
equations for each length $L$. A comparison between ${\cal N}(L)$ and ${\cal N}_{CFT}(L)$    
for $\gamma=\pi/2$ and $\gamma = \pi/3$ is exhibited in Tables~\ref{t1} and~\ref{t2}.

\begin{table}[hbt]
\centering
\begin{tabular}{|c||c|c|c|}
\hline
$L$ & ${\cal N}(L)$  & ${\cal N}_{CFT}(L)$        \\ \hline \hline
    4  & \,0.457106781187\, & \,0.446378653269\, \\ \hline
    8  & \,0.366669830087\, & \,0.366041268056\, \\ \hline
   16  & \,0.345995599194\, & \,0.345956921753\, \\ \hline
   32  & \,0.340938243195\, & \,0.340935835178\,   \\ \hline
   64  & \,0.339680713890\, & \,0.339680563534\, \\ \hline
  128  & \,0.339366755018\, & \,0.339366745623\, \\ \hline
  256  & \,0.339288291732\, & \,0.339288291145\, \\ \hline
  512  & \,0.339268677562\, & \,0.339268677525\, \\ \hline
 1024  & \,0.339263774123\, & \,0.339263774121\, \\ \hline
\end{tabular}
\caption[table1]{Comparison between ${\cal N}(L)$ 
and ${\cal N}_{CFT}(L)$ for $\gamma=\pi/2$ (the XX model). 
For an infinite chain, we have that negativity is given 
by ${\cal N}(\infty)=0.339262139652$.}
\label{t1}
\end{table}

\begin{table}[hbt]
\centering
\begin{tabular}{|c||c|c|c|}
\hline
$L$ & ${\cal N}(L)$  & ${\cal N}_{CFT}(L)$        \\ \hline \hline
    4  & \,0.489830037812\, & \,0.478556230132\, \\ \hline
    8  & \,0.401639244141\, & \,0.400889057533\, \\ \hline
   16  & \,0.381525197365\, & \,0.381472264383\, \\ \hline
   32  & \,0.376621871264\, & \,0.376618066096\, \\ \hline
   64  & \,0.375404791436\, & \,0.375404516524\, \\ \hline
  128  & \,0.375101148980\, & \,0.375101129131\, \\ \hline
  256  & \,0.375025283711\, & \,0.375025282283\, \\ \hline
  512  & \,0.375006320673\, & \,0.375006320571\, \\ \hline
 1024  & \,0.375001580150\, & \,0.375001580143\, \\ \hline
\end{tabular}
\caption[table1]{Comparison between ${\cal N}(L)$ 
and ${\cal N}_{CFT}(L)$ for $\gamma=\pi/3$. For an infinite chain, we have 
that negativity is given by ${\cal N}(\infty)=3/8=0.375$.}
\label{t2}
\end{table}

\subsection{Twisted boundary conditions}
We can also use the results obtained for PBC to investigate the finite size corrections to 
the negativity with more general boundary conditions. We will consider here the so-called 
twisted boundary conditions (TBC), which can be achieved as the effect of 
a magnetic flux through a spin ring~\cite{Byers:61}. Remarkably, it has recently been shown that TBC 
may improve multi-party quantum communication via spin chains~\cite{Bose:05}. In order to consider 
TBC, it is convenient to rewrite the Hamiltonian in Eq.~(\ref{HXXZ}) (with $J=1$) in the 
following form
\begin{equation}
H_{XXZ} = -\frac{1}{2} \sum_{i=1}^{L} \left[ 2 \left( \sigma^+_i \sigma^-_{i+1} + 
\sigma^-_i \sigma^+_{i+1} \right) + \Delta \sigma^z_i \sigma^z_{i+1} \right], 
\label{hfitbc}
\end{equation}
where $\sigma^{\pm}_j = (\sigma^{x}_j \pm i \sigma^{y}_j)/2$ 
and $\sigma^{\pm}_{L+1} = e^{\pm i\Phi} \sigma^{\pm}_{1}$ ($0 \le \Phi < 2 \pi$), with $\Phi$ denoting a phase. 
The quantum chain given by Eq.~(\ref{hfitbc}) is solvable by the Bethe ansatz~\cite{Alcaraz:88}. 
In presence of TBC, Eq.~(\ref{energy}) still holds, but 
with an effective central $\hat{c}(\Phi)$~\cite{Alcaraz:88}, which is given by 
\begin{equation}
\hat{c}(\Phi) = 1 - \frac{3 \Phi^2}{2 \pi \left(\pi - \gamma\right)} ,
\label{effcc}
\end{equation}
with $\gamma$ defined as in Eq.~(\ref{gamadef}). Let us take the following canonical transformations~\cite{Alcaraz:90}
\begin{equation}
\widetilde{\sigma}^{\pm}_j = e^{\mp i \Phi j/L} \sigma^{\pm}_j \,\,\, , \,\,\, 
\widetilde{\sigma}^{z}_j = \sigma^{z}_j \,\,\,\, (j=1,\ldots,L)\, .
\label{nv}
\end{equation} 
In terms of this new set of operators, the original chain with TBC is now given by the periodic chain
\begin{eqnarray}
H_{XXZ} &=& -\frac{1}{2} \sum_{j=1}^{L} \left[ e^{-i \Phi/L} \widetilde{\sigma}^+_j 
\widetilde{\sigma}^-_{j+1} + e^{i \Phi/L} \widetilde{\sigma}^-_j \widetilde{\sigma}^+_{j+1} \right. \nonumber \\
&+& \left. \Delta \widetilde{\sigma}^z_j \widetilde{\sigma}^z_{j+1} \right] \, ,
\end{eqnarray} 
where $\widetilde{\sigma}^{\pm}_{L+1} = \widetilde{\sigma}^{\pm}_1$. 
Defining the operators $\widetilde{\sigma}^{x}_j$ and $\widetilde{\sigma}^{y}_j$ through 
$\widetilde{\sigma}^{\pm}_j = (\widetilde{\sigma}^{x}_j \pm i \widetilde{\sigma}^{y}_j)/2$, 
the Hamiltonian can be put in the form
\begin{eqnarray}
H_{XXZ} = -\frac{1}{2} \sum_{j=1}^{L} \left[ \cos\left(\frac{\Phi}{L} \right) \frac{(\widetilde{\sigma}^x_j 
\widetilde{\sigma}^x_{j+1} + \widetilde{\sigma}^y_j \widetilde{\sigma}^y_{j+1})}{2} \right. 
\nonumber \\
- \left. \sin \left(\frac{\Phi}{L}\right) \frac{(\widetilde{\sigma}^x_j 
\widetilde{\sigma}^y_{j+1} - \widetilde{\sigma}^y_j \widetilde{\sigma}^x_{j+1})}{2}
+ \Delta \widetilde{\sigma}^z_j \widetilde{\sigma}^z_{j+1} \right].
\label{htbc}
\end{eqnarray}
Note that the Hamiltonian in Eq.~(\ref{htbc}) is both U(1) invariant ($[H,\sum_j \widetilde{\sigma}^{z}_j]=0$) 
and translationally invariant ($H_{XXZ}$ exhibts PBC in terms 
of the set $\{\widetilde{\sigma}^{\pm}_j,\widetilde{\sigma}^{z}_j\}$). Therefore,  
the two-spin reduced density matrix keeps the form given in Eq.~(\ref{rij}), with 
the correlation functions $G^{ij}_{\alpha\beta}$ replaced by 
$\widetilde{G}^{ij}_{\alpha\beta}=\langle \widetilde{\sigma}_\alpha^i 
\widetilde{\sigma}_\beta^{j} \rangle$. Then, the negativity for nearest neighbor spins governed by Hamiltonian~(\ref{htbc}) can be 
computed similarly as before. By using that $G_z=0$ (ground state) and $|G^{i,i+1}_{zz}| \le 1$ 
we obtain
\begin{equation}
{\cal N} = 2 \max \left( 0 , |z| - a \right), 
\label{negtbc}
\end{equation}
where $a = (1+\widetilde{G}_{zz})/4$ and $z = (\widetilde{G}_{\parallel} + i  
\widetilde{G}_{\perp})/4$, with $\widetilde{G}_{zz} = \widetilde{G}^{i,i+1}_{zz}$, 
$\widetilde{G}_{\parallel} = \widetilde{G}^{i,i+1}_{xx} + \widetilde{G}^{i,i+1}_{yy}$, 
and $\widetilde{G}_{\perp} = \widetilde{G}^{i,i+1}_{xy} - \widetilde{G}^{i,i+1}_{yx}$ 
($\forall \, i$). In order to write entanglement in terms of the derivatives of the 
energy density, it is convenient to define $\overline{H}_{XXZ} = H_{XXZ} / \cos(\Phi/L)$. Then 
\begin{equation}
\overline{\varepsilon} = - \frac{1}{2} \left( \widetilde{G}_{\parallel} - \eta 
\widetilde{G}_{\perp} + \overline{\Delta} \widetilde{G}_{zz} \right) , 
\label{ednh}
\end{equation}
where $\overline{\varepsilon} = \langle \overline{H}_{XXZ} \rangle / L$, 
$\eta = \tan (\Phi/L)$, and $\overline{\Delta} = \Delta / \cos(\Phi/L)$. From Eq.~(\ref{ednh}) 
we get
\begin{eqnarray}
\widetilde{G}_{zz} &=& - 2 \frac{\partial \overline{\varepsilon}}{\partial \overline{\Delta}} \, , \nonumber \\
\widetilde{G}_{\perp} &=& 2 \frac{\partial \overline{\varepsilon}}{\partial \eta} \, , \nonumber \\
\widetilde{G}_{\parallel} &=& 2 \left( - \overline{\varepsilon} + \eta \frac{\partial \overline{\varepsilon}}{\partial \eta} + \overline{\Delta}\frac{\partial \overline{\varepsilon}}{\partial \overline{\Delta}} \right) \, .
\label{enderiv}
\end{eqnarray}
Therefore, the contribution $(|z|-a)$ for expression for the negativity in Eq.~(\ref{negtbc}) reads
\begin{eqnarray}
|z| - a &=& \frac{1}{2} \left( \sqrt{\left(- \overline{\varepsilon} + \eta \frac{\partial \overline{\varepsilon}}{\partial \eta} + \overline{\Delta}\frac{\partial \overline{\varepsilon}}{\partial \overline{\Delta}} \right)^2 + 
\left(\frac{\partial \overline{\varepsilon}}{\partial \eta} \right)^2} \right. \nonumber \\ 
&+& \left. \frac{\partial \overline{\varepsilon}}{\partial \overline{\Delta}} - \frac{1}{2} \right) .
\label{fe}
\end{eqnarray}
In order to obtain the results in terms of $\Phi$ and $\Delta$, we make use of the 
expressions
\begin{eqnarray}
\frac{\partial \overline{\varepsilon}}{\partial \eta} &=& \cos\left(\frac{\Phi}{L}\right) 
\left( L \frac{\partial\varepsilon}{\partial\Phi} + \tan\left(\frac{\Phi}{L}\right) \varepsilon \right) \, , \label{oed1}   \\
\frac{\partial \overline{\varepsilon}}{\partial \overline{\Delta}} &=&
\frac{\partial \varepsilon}{\partial \Delta} \, .
\label{oed2}
\end{eqnarray}
Hence, finite size corrections to entanglement can be found now by using Eq.~(\ref{energy}) [replacing 
the central charge $c$ by the effective central charge $\hat{c}(\Phi)$ as in Eq.~(\ref{effcc})] into Eq.~(\ref{fe}). 
Examples comparing the negativity ${\cal N}_{CFT}(L)$ for nearest neighbors up to $o(L^{-2})$ and the exact value of the negativity 
${\cal N}(L)$ (obtained through the numerical solution of the Bethe ansatz equations derived in Ref.~\cite{Alcaraz:88}) are 
exhibited in Tables~\ref{t3} and~\ref{t4} below. 

\begin{table}[hbt]
\centering
\begin{tabular}{|c||c|c|c|}
\hline
$L$ & ${\cal N}(L)$  & ${\cal N}_{CFT}(L)$        \\ \hline \hline
    4  & \,0.406774810601\, & \,0.446378653269\, \\ \hline
    8  & \,0.354315234931\, & \,0.366041268056\, \\ \hline
   16  & \,0.342922395530\, & \,0.345956921753\, \\ \hline
   32  & \,0.340170924101\, & \,0.340935835178\, \\ \hline
   64  & \,0.339488945731\, & \,0.339680563534\, \\ \hline
  128  & \,0.339318816833\, & \,0.339366745623\, \\ \hline
  256  & \,0.339276307427\, & \,0.339288291145\, \\ \hline
  512  & \,0.339265681501\, & \,0.339268677525\, \\ \hline
 1024  & \,0.339263025109\, & \,0.339263774121\, \\ \hline
\end{tabular}
\caption[table1]{Comparison between the exact evaluation of ${\cal N}(L)$ 
and the approximate expression ${\cal N}_{CFT}(L)$ (up to order $L^{-2}$)) for 
$\gamma = \pi /2$ and $\Phi = \pi/2$.}
\label{t3}
\end{table}
\begin{table}[hbt]
\centering
\begin{tabular}{|c||c|c|c|}
\hline
$L$ & ${\cal N}(L)$  & ${\cal N}_{CFT}(L)$        \\ \hline \hline
    4  & \,0.400000000000\, & \,0.452230707893\, \\ \hline
    8  & \,0.381121448251\, & \,0.394307676973\, \\ \hline
   16  & \,0.376577662094\, & \,0.379826919243\, \\ \hline
   32  & \,0.375405200439\, & \,0.376206729811\, \\ \hline
   64  & \,0.375102994373\, & \,0.375301682453\, \\ \hline
  128  & \,0.375025983614\, & \,0.375075420613\, \\ \hline
  256  & \,0.375006526789\, & \,0.375018855153\, \\ \hline
  512  & \,0.375001635654\, & \,0.375004713788\, \\ \hline
 1024  & \,0.375000409414\, & \,0.375001178447\, \\ \hline
\end{tabular}
\caption[table1]{Comparison between the exact evaluation of ${\cal N}(L)$ 
and the approximate expression ${\cal N}_{CFT}(L)$ (up to order $L^{-2}$)) for 
for $\gamma = \pi /3$ and $\Phi = 2\pi/3$.}
\label{t4}
\end{table}

Note from Tables~\ref{t1} and~\ref{t3} that, for $\gamma = \pi/2$, ${\cal N}_{CFT}(L)$ gives the same result either for $\Phi=0$ (PBC) or 
$\Phi=\pi/2$, which is an indication that TBC should not affect the negativity (up to $o(L^{-2})$) in the case of the XX model. 
Indeed, this can be analytically proved. In this case, the anisotropy is $\Delta=0$, 
which implies that $\gamma=\pi/2$ and $\xi=2$. Then, from Eqs.~(\ref{energy}) and~(\ref{effcc}), we have 
\begin{eqnarray}
\varepsilon(L) &=& \varepsilon_\infty-\frac{\pi\,\hat{c}(\Phi)}{3}L^{-2}+o(L^{-2}), \nonumber \\
\frac{\partial\varepsilon}{\partial\Phi} &=& \frac{2}{\pi} \frac{\Phi}{L^2} +o(L^{-2}),\nonumber \\
\frac{\partial\varepsilon}{\partial\Delta} &=& \left.\frac{\partial\varepsilon_\infty}{\partial\Delta}\right|_{\Delta=0} + \frac{2}{3L^2}+o(L^{-2}).
\end{eqnarray}
By inserting the above equations into Eqs.~(\ref{oed1}) and~(\ref{oed2}), it can be shown that the negativity as given by Eq.~(\ref{negtbc}) gets
\begin{equation}
{\cal N}_{CFT} = \left| \varepsilon_{(\Phi=0)} \right| + \left.\frac{\partial \varepsilon_\infty}{\partial \Delta}\right|_{\Delta= 0} + \frac{2}{3L^2} -\frac{1}{2} +o(L^{-2})  , 
\label{negTBCXX}
\end{equation}
where $ |\varepsilon_{(\Phi=0)}| = \sqrt{\varepsilon_\infty^2 - 2 \varepsilon_\infty \pi \xi / (6L^2)}$. 
Hence, Eq.~(\ref{negTBCXX}) implies that the negativity for the XX model with TBC is not affected by the phase 
$\Phi$ up to order $1/L^2$.

\subsection{Excited states}

\begin{figure}[ht]
\centering {\includegraphics[angle=0,scale=0.36]{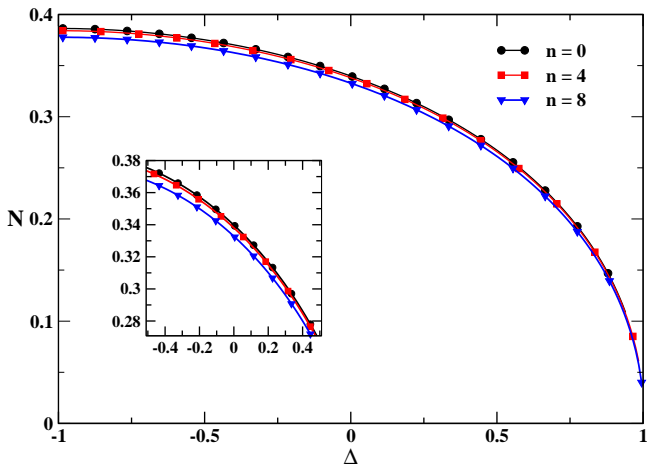}}
\caption{(color online) Negativity for minimum energy states as a function of the anisotropy $\Delta$ for $L=256$ sites. 
Note that the curves are nearly the same, indicating a unique entanglement pattern in the critical region. 
Inset: Negativity as a function of $\Delta$ in a larger zoom scale.}
\label{f2}
\end{figure}

Let us consider now the structure of the negativity for the excited states in the XXZ model with PBC. 
The $U(1)$ and translation symmetries allow us the decomposition of the associated eigenspace of $H_{XXZ}$ into 
disjoint sectors (fixed magnetization and momentum) labelled by the quantum numbers $r=0,1,2,\ldots,L$ and 
$p=1,2,\ldots,L-1$, which give the number of spins up in the $\sigma_z$ basis and the eigenvalue of the momentum $P=(2\pi/L)p$, 
respectively.  
An exact evaluation of the negativity for nearest neighbor spins can be performed from Eq.~(\ref{negXXZ_gen}) by taking a 
non-vanishing magnetization density $G_z$ and by using Eq.~(\ref{correl}), where the energy of the excited state is obtained 
through the solution of the Bethe ansatz equations. This is illustrated in Fig.~\ref{f2}, where we 
plot the negativity between nearest neighbors in a chain of length $L=256$ sites for minimum energy 
states with zero momentum in several magnetization sectors. 
These states have anomalous dimensions $x_n$ given by~\cite{Alcaraz:88}
\begin{equation}
x_n = n^2 \frac{(\pi - \gamma)}{2\pi},
\label{ad}
\end{equation}
where $n = L/2 - r$ and $j+j^\prime=p=0$ in Eq.~(\ref{energy_ex}). Remarkably, note that the negativities for the minimum energy states plotted are nearly the same, indicating a unique entanglement pattern in the critical region. Indeed, this is a more general 
result, which holds also for other excited states. For instance, let us take the so-called marginal state~\cite{Alcaraz:88}, which is a state that will be taken in the sector $n=0$ 
with anomalous dimension $x=2$ (independently of $\gamma$) and $j,j^\prime=0$. Exact computation in Table~\ref{t5} below shows that its 
negativity is also close to the values found in Fig.~\ref{f2}. 
\begin{table}[hbt]
\centering
\begin{tabular}{|c||c|c|c|}
\hline
$\Delta$ & Marginal State  & Ground State  \\ \hline \hline
 0.505023772863  & \,0.265447369819\, & \,0.266151418398\, \\ \hline
 0.205023772863  & \,0.315358910123\, & \,0.316005319520\, \\ \hline
 0.005023772863  & \,0.338660739066\, & \,0.339288291732\, \\ \hline
 -0.204976227137 & \,0.357090720201\, & \,0.357706303640\, \\ \hline
 -0.504976227137 & \,0.374867541783\, & \,0.375473489099\, \\ \hline
\end{tabular}
\caption[table1]{Comparison between ${\cal N}(\Delta)$ for the marginal state and  
and ${\cal N}(\Delta)$ for the ground state in a chain with $L=256$ sites.}
\label{t5}
\end{table}
Indeed, we can show that entanglement in the critical region of the XXZ chain will exhibit a unique pattern for all states accessible via the 
CFT associated with the model. 
As discussed in Section~\ref{enerCFT}, each primary operator of the theory corresponds to a tower of states 
with energies given by Eq.~(\ref{energy_ex}). All these states in the towers will have energies which differ at order $L^{-2}$ [see 
Eqs.~(\ref{energy}) and~(\ref{energy_ex})]. According to Eq.~(\ref{negXXZ_gen}), such a difference is also reflected in the negativity of 
nearest neighbor spins, which explains the behavior displayed in both Fig.~\ref{f2} and Table~\ref{t5}. This can explicitly be shown by 
inserting Eq.~(\ref{energy_ex}) into Eq.~(\ref{negXXZ_gen}). As an illustration, we take the minimum energy states with zero momentum in a given 
magnetization sector labelled by $n$. For this case, the negativity can be evaluated as
\begin{eqnarray}
&& {\cal N}_{n}(L) = {\cal N}_n^\infty + \frac{\left| G_\infty \right|^{-1}}{6 \gamma L^2} \left[ \frac{}{} 3 \gamma n^2  + \pi \, \sin\gamma \, G_\infty  z_n \right. \nonumber \\
&&\left. + \pi \left(\Delta \, G_\infty + \left| G_\infty \right|\right) \left(  
\frac{w_n}{\gamma} - z_n \cot \gamma   \right) \right] + o(L^{-2}), \nonumber
\label{negn}
\end{eqnarray}
where 
\begin{eqnarray}
&& w_n = \pi ( c - 6n^2 ), \,\,\, {\cal N}_n^\infty = | G_\infty | + \partial\varepsilon_n^\infty / \partial \Delta -1/2 , \nonumber \\
&&  z_n = \left( \pi c - 6 n^2 (\pi - \gamma)\right), \,\, G_\infty = - \varepsilon_n^\infty + \Delta \,\partial\varepsilon_n^\infty / \partial \Delta. \nonumber 
\end{eqnarray}
with $\varepsilon_n^\infty$ denoting the energy density of the excited state as $L \rightarrow \infty$. 
Note that this unique pattern of entanglement, which 
has been explicitly derived here, is in agreement with the general discussion of Sec.~\ref{ecft}. This is indeed exhibited in Eq.~(\ref{mcft}). 
Naturally, similar expressions can be obtained for excited states higher than the minimum energy states. 

\section{Conclusion} 

In conclusion, we have investigated the computation of finite size corrections to 
entanglement in quantum critical systems. These corrections were shown to depend on 
the central charge of the model as well as the anomalous dimensions of the primary 
operators of the theory. 
Our approach has naturally arisen as a 
general consequence of the application of CFT and DFT methods in critical theories. 
This framework has been illustrated in the XXZ model, where we have shown that: 
(i) entanglement in spin chains with arbitrary finite sizes can be analytically 
computed up to order $o(L^{-2})$ with no need of solving the Bethe 
ansatz equations for each length $L$; (ii) Conformal towers of excited states displays a 
unique pattern of entanglement in the critical region. 
Indeed, we have been able to provide a general argument according to which this unique pattern of 
entanglement should appear in all conformally invariant models.  
Further examples in higher dimensional lattices and higher spin systems 
are left for future investigation.

\subsection*{Acknowledgments}

This work was supported by the Brazilian agencies MCT/CNPq (F.C.A, M.S.S.),  
FAPESP (F.C.A.), and FAPERJ (M.S.S.).


\begin{thebibliography}{9}

\bibitem{Osterloh:02} A. Osterloh, L. Amico, G. Falci, and R. Fazio, Nature {\bf 416}, 608 (2002).

\bibitem{Nielsen:02} T. J. Osborne and M. A. Nielsen, Phys. Rev. A {\bf 66}, 032110 (2002).

\bibitem{Vidal:03} G. Vidal, J. I. Latorre, E. Rico, and A. Kitaev, Phys. Rev. Lett. {\bf 90}, 227902 (2003).

\bibitem{Amico:08} L. Amico, R. Fazio, A. Osterloh, and V. Vedral, Rev. Mod.
Phys. {\bf 80}, 517 (2008).

\bibitem{Korepin:04} V. E. Korepin, Phys. Rev. Lett. {\bf 92}, 096402 (2004).

\bibitem{Calabrese:04} P. Calabrese and J. Cardy, J. Stat. Mech. {\bf 0406}, 002 (2004).

\bibitem{Keating:05} J. P. Keating and F. Mezzadri, Phys. Rev. Lett. {\bf 94}, 050501 (2005).

\bibitem{Laflorencie:05} N. Laflorencie, E. S. S{\o}rensen, M.-S. Chang, and I. Affleck, 
Phys. Rev. Lett. {\bf 96}, 100603 (2006). 

\bibitem{Refael:04} G. Refael and J. E. Moore, Phys. Rev. Lett. {\bf 93}, 260602 (2004).

\bibitem{Saguia:07} A. Saguia, M. S. Sarandy, B. Boechat, and M. A. Continentino, 
Phys. Rev. A {\bf 75}, 052329 (2007).

\bibitem{Lin:07} Y.-C. Lin, F. Igl\'oi, and H. Rieger, Phys. Rev. Lett. {\bf 99}, 147202 (2007).

\bibitem{Hur:07} K. Le Hur, P. Doucet-Beaupr\'e, and W. Hofstetter, 
Phys. Rev. Lett. {\bf 99}, 126801 (2007).

\bibitem{Fuehringer:08} M. Fuehringer, S. Rachel, R. Thomale, M. Greiter, and P.
Schmitteckert, e-print arXiv:0806.2563 (2008).

\bibitem{Franca:08} V. V. Fran\c{c}a and K. Capelle, Phys. Rev. Lett. {\bf 100}, 070403 (2008).

\bibitem{Blote:86} H. W. J. Bl\"ote, J. L. Cardy, and M. P. Nightingale, Phys. Rev. 
Lett. {\bf 56}, 742 (1986).

\bibitem{Affleck:86} I. Affleck, Phys. Rev. Lett. {\bf 56}, 746 (1986).

\bibitem{Cardy:86} J. L. Cardy, Nucl. Phys. B {\bf 270 [FS16]}, 186 (1986). 

\bibitem{Schonhammer:95} K. Sch\"{o}nhammer, O. Gunnarsson, and R. M. Noack,
Phys. Rev. B \textbf{52}, 2504 (1995).

\bibitem{Wu:05} L.-A. Wu, M. S. Sarandy, D. A. Lidar, and L. J. Sham, 
Phys. Rev. A {\bf 74}, 052335 (2006).

\bibitem{Wootters:98} W. K. Wootters, Phys. Rev. Lett. {\bf 80}, 2245 (1998).

\bibitem{Vidal:02} G. Vidal and R. F. Werner, Phys. Rev. A {\bf 65},  032314  (2002).

\bibitem{Bose:02} I. Bose and E. Chattopadhyay, Phys. Rev. A {\bf 66}, 062320 (2002).

\bibitem{Alcaraz:04} F. C. Alcaraz, A. Saguia, and M. S. Sarandy, Phys. Rev. A {\bf 70}, 032333 (2004).

\bibitem{Wu:04} L.-A. Wu, M. S. Sarandy, and D. A. Lidar, Phys. Rev. Lett. {\bf 93}, 250404 (2004).

\bibitem{Yang:05} M.-F. Yang, Phys. Rev. A \textbf{71}, 030302(R) (2005).

\bibitem{Gehlen:86} G. von Gehlen, V. Rittenberg, and H. Ruegg, J. Phys. A {\bf 19}, 107 (1986). 

\bibitem{Alcaraz:87} F. C. Alcaraz, M. N. Barber, and M. T. Batchelor, Phys. Rev. Lett. 
{\bf 58}, 771 (1987). 

\bibitem{Alcaraz:88} F. C. Alcaraz, M. N. Barber, and M. T. Batchelor, Ann. Phys. (N.Y.) 
{\bf 182}, 280 (1988). 

\bibitem{Hohenberg:64} P. Hohenberg and W. Kohn, Phys. Rev. \textbf{136}, B864
(1964).

\bibitem{Kohn:65} W. Kohn and L. J. Sham, Phys. Rev. \textbf{140}, A1133 (1965).

\bibitem{Capelle:01} K. Capelle and G. Vignale, Phys. Rev. Lett. {\bf 86}, 5546 (2001).

\bibitem{Hellmann:37} H. Hellmann, 
{\it Die Einf\"uhrung in die Quantenchemie} (Deuticke, Leipzig, 1937).

\bibitem{Feynman:39} R. P. Feynman, Phys. Rev. {\bf 56}, 340 (1939).

\bibitem{Sagdeev:89} I. R. Sagdeev and A. B. Zamolodchikov, Mod. Phys. Lett. B {\bf 3}, 1375 (1989).

\bibitem{Hamer:86} C. J. Hamer, J. Phys. A {\bf 19}, 3335 (1986).

\bibitem{Yang:66} C. N. Yang and C. P. Yang, Phys. Rev. {\bf 150}, 321 (1966); {\it ibid.} 
{\bf 150}, 327 (1966).

\bibitem{Byers:61} N. Byers and C. N. Yang, Phys. Rev. Lett. {\bf 7}, 46 (1961).

\bibitem{Bose:05} S. Bose, B.-Q. Jin, and V. E. Korepin, Phys. Rev. A {\bf 72}, 022345 (2005).

\bibitem{Alcaraz:90} F. C. Alcaraz and W. F. Wreszinski, J. Stat. Phys. {\bf 58}, 45 (1990).

\end{thebibliography}
\end{document}